\documentclass[12pt]{article}
\usepackage{color,amssymb,amsmath,hyperref}
\usepackage{empheq}
\usepackage{url}

\definecolor{rougef}{rgb}{0.56,0,0}
\definecolor{vertf}{rgb}{0,0.5,0}
\definecolor{bleuf}{rgb}{0,0,0.8}

 \csname
@addtoreset\endcsname{equation}{section}

\def\a{\alpha}

\def\b{\beta}   
\def\c{\gamma} 
\def\C{\Gamma}
\def\d{\delta} 

\def\e{\epsilon}

\def\k{\kappa}
\def\l{\lambda}

\def\s{\sigma}

\def\del{\partial}

\def\nn{\nonumber}
\newcommand{\eq}[1]{(\ref{#1})}

\newcommand{\w}[1]{\\[0.#1cm]}
 
\def\be{\begin{equation}}
\def\ee{\end{equation}}
\def\bea{\begin{eqnarray}}
\def\eea{\end{eqnarray}}
\def\ba{\begin{array}}
\def\ea{\end{array}}

\def\ed{\end{document}}

\def\Real{{\mathbb R}}
\def\Comp{{\mathbb C}}

\def\de{\partial}
\def\yt{\tilde{y}} \def\zt{\tilde{z}}
\newcommand{\abs}[1]{\lvert#1\rvert}

\begin{document}

\begin{titlepage}

\setcounter{page}{1}

\begin{center}

\hfill MI-TH-1543

\vskip 1cm


{\Large \bf  An Action for Matter Coupled Higher Spin
Gravity\\[5pt] in Three Dimensions}


\vspace{12pt}

Roberto Bonezzi\,$^1$, Nicolas Boulanger\,$^1$, 
Ergin Sezgin\,$^2$ and Per Sundell\,$^{3}$\\

\vskip 25pt

{\em $^1$ \hskip -.1truecm Physique th\'eorique et math\'ematique, 
Universit\'e de Mons -- UMONS \\
20, Place du Parc, 7000 Mons, Belgium\vskip 5pt }

\vskip 10pt

{\em $^2$ \hskip -.1truecm George and Cynthia Woods Mitchell Institute for Fundamental
Physics and Astronomy \\ Texas A\& M University, College 
Station, TX 77843, USA }

\vskip 10pt
 
{\em $^3$ \hskip -.1truecm Departamento de Ciencias F\'isicas, Universidad Andres Bello, 
\\ Republica 220, Santiago de Chile}

\end{center}

\vskip 15pt

\begin{center}

{emails: \small{{\tt roberto.bonezzi@umons.ac.be},  
{\tt nicolas.boulanger@umons.ac.be}, 
{\tt sezgin@tamu.edu}, 
{\tt per.sundell@unab.cl}}}
\end{center}

\vskip .5cm

\begin{center} {\bf ABSTRACT}\\[1ex]
\end{center}

We propose a covariant Hamiltonian action for 
the Prokushkin and Vasiliev's matter 
coupled higher spin gravity in three dimensions.
The action is formulated on ${\cal X}_4 \times {\cal Z}_2$ 
where ${\cal X}_4$ is an open manifold whose boundary 
contains spacetime and ${\cal Z}_2$ is a noncommutative 
twistor space.
We examine various consistent truncations to models of
BF type in 
${\cal X}_4$ and ${\cal Z}_2$ with B$^2$ terms and central 
elements.
They are obtained by integrating out the matter fields in 
the presence of a vacuum expectation value $\nu\in \Real$ for the 
zero-form master field.
For $\nu=0$, we obtain a model on ${\cal X}_4$ 
containing Blencowe's action and a model on ${\cal Z}_2$ 
containing the Prokushkin--Segal--Vasiliev action.
For generic $\nu$ (including $\nu=0$), we propose
an alternative model on ${\cal X}_4$ with gauge 
fields in the Weyl algebra of Wigner's deformed
oscillator algebra and Lagrange multipliers in the algebra
of operators acting in the Fock representation space 
of the deformed oscillators.

\end{titlepage}

\newpage

\tableofcontents

\vspace{1 cm}


\section{Introduction}


Three-dimensional gravity with negative cosmological constant, 
as defined by the $sl(2,\mathbb{R})\oplus sl(2,\mathbb{R})$ 
Chern--Simons (CS) action of \cite{Achucarro:1987vz,Witten:1988hc}, 
provides a rich framework for testing various aspects of quantum 
gravity in a setting that is simpler than in higher dimensions, 
yet nontrivial. 
Although AdS$_3$ gravities are topological, they admit black holes 
\cite{Banados:1992wn} and possess moduli spaces at conformal infinity 
governed by infinite-dimensional conformal symmetry algebras 
\cite{Brown:1986nw,Strominger:1997eq}; see \cite{Carlip:2005zn} 
for a review. 
\vspace{.3cm}

As for higher spin gravities, these are simpler as well in three 
dimensions, where the massless higher spin fields are topological,
and hence the spectrum requirements on the gauge algebras 
simplify considerably.
Topological higher spin gravities based on the 
principal embedding of $sl(2,\mathbb{R})$ into
$sl(N,\mathbb{R})$ were shown in \cite{Henneaux:2010xg,Campoleoni:2010zq}
to have asymptotic $W_N$ symmetries. 
In \cite{Gaberdiel:2010pz} holographic correspondences 
were conjectured for higher spin gravities with infinite-dimensional
gauge algebras $hs(2)\oplus hs(2)$ and their deformation
$hs(\lambda)\oplus hs(\lambda)$ coupled to (complex) bulk scalars.
\vspace{.3cm}

In the above works, the (classical) gauge sector is assumed to
be described by various CS generalisations 
\cite{Blencowe:1988gj,Bergshoeff:1989ns,Fujisawa:2013ima} 
of the Achucarro--Townsend supergravity Lagrangian 
\cite{Achucarro:1987vz}.
On the other hand, one class of matter coupled 
higher spin gravities is described on-shell by the 
Prokushkin--Vasiliev (PV) equations \cite{Prokushkin:1998bq},
and off-shell by the Prokushkin--Segal--Vasiliev (PSV) action
principle in twistor space \cite{Prokushkin:1999gc}. 
However, as spacetime is absent in the latter, its relation 
to the CS formulations has remained unclear.
Moreover, there exists a second class of matter coupled 
higher spin gravities based on action principles in
three dimensions with an extra dynamical two-form
\cite{Fujisawa:2013ima}, whose relation to the PV system
is unclear.
\vspace{.3cm}

In this paper, we provide the PV system with an action 
principle of covariant Hamiltonian type on a six-manifold
given by the direct product of a closed twistor space 
${\cal Z}_2$ and an open four-manifold ${\cal X}_4$ 
whose boundary ${\cal X}_3$ contains spacetime ${\cal M}_3$.
When subjected to the variational principle combined 
with natural boundary conditions, the action yields 
the PV equations on its five-dimensional boundary 
${\cal X}_3 \times {\cal Z}_2$.
The action is constructed such that upon
integrating out the matter fields in the presence of
an expectation value $\nu$ for the PV zero-form, 
the effective action for the gauge fields can 
be consistently truncated to models of BF type on 
${\cal X}_4$. 
These model contain B squared terms\footnote{
The B field of the BF-like models originate from 
the even Lagrange multiplier in the covariant 
Hamiltonian action, denoted by $T$ in \eq{ma}.}
containing the standard symplectic structure of 
three-dimensional Fronsdal fields.
\vspace{.3cm}

For $\nu=0$, we obtain a model on ${\cal X}_4$ 
containing Blencowe's action, as well as a model on ${\cal Z}_2$ 
containing the Prokushkin--Segal--Vasiliev action.
In these models, the dual spaces that contain the gauge fields and 
Lagrange multipliers are isomorphic.
For generic $\nu$ (including $\nu=0$), we shall also consider
a model on ${\cal X}_4$ in which the aforementioned
two spaces are not isomorphic. 
However, its existence depends on the finiteness of the trace of the 
vacuum-to-vacuum projector of a deformed oscillator 
induced from six dimensions.
If existing, such a model would provide an alternative to 
the BF-like Blencowe model based on Vasiliev's 
supertrace \cite{Vasiliev:1989re}, whose six-dimensional 
origin remains unclear.
\vspace{.3cm}

The master action to be constructed here
is analog of the one for the four-dimensional 
Vasiliev system found in \cite{Boulanger:2011dd}.
In particular, it does \emph{not} extend the 
closed and central two-form of the PV system
into a dynamical field off-shell.
The inclusion of a dynamical two-form in an action 
that is an analog of that for four-dimensional
models given recently in \cite{Boulanger:2015kfa} and which makes 
contact with both the PV system as well as the
action proposed in \cite{Fujisawa:2013ima} will be treated elsewhere.
\vspace{.3cm}

The paper is structured as follows: In Section \ref{sec:geomPV},
we cast the PV system as a differential algebra on the
direct product of twistor space and spacetime.
After this preparation, we propose a covariant Hamiltonian
action on ${\cal X}_4\times {\cal Z}_2$, with a term quadratic in
Lagrange multipliers, in Section 3.
In Section \ref{Sec:Blencowe}, we examine the 
consistent truncations to the BF-like version of the 
Blencowe model on ${\cal X}_4$ and the PSV action on ${\cal Z}_2$. 
We summarize our results and provide an outlook in the Conclusions 
in Section \ref{sec:Ccl}. 
In Appendix A, we review the mass deformation.
In Appendix B, we present the proposed $\nu$-deformed 
BF-like model on ${\cal X}_4$.

%
\section{Prokushkin--Vasiliev models}
\label{sec:geomPV}

In this section, we rewrite the PV equations \cite{Prokushkin:1998bq} 
as a differential algebra generated by master fields on
a noncommutative manifold valued in an associative algebra, or
equivalently, as an associative bundle with fusion rules.
In particular, we shall identify the minimal bosonic model and its
massive deformation. For a recent, in-depth treatment of the weak-field
perturbative analysis of PV systems to first nontrivial order 
in interactions, see \cite{Kessel:2015kna}.

\subsection{Differential algebra} 

The master fields are
\begin{equation}
A=dx^\mu W_\mu(x,z|y;\Gamma_i)+dz^\a V_\a(x,z|y;\Gamma_i)\;,\quad B=B(x,z|y;\Gamma_i)\;,
\end{equation}
defined locally on the direct product ${\cal M}_3\times {\cal Z}_2$ 
of a commutative three-dimensional real manifold ${\cal M}_3$ with 
coordinates $x^\mu$, $\mu=1,2,3$, and a non-commutative two-dimensional 
real manifold ${\cal Z}_2$ with coordinates $z^\a$, $\a=1,2$.
The fields are valued in an associative algebra generated by
a real oscillator $y^\a$, $\a=1,2$, coordinatizing an internal
noncommutative manifold ${\cal Y}_2$, and a set of elements 
$\Gamma_i$, $i=1,\dots, N$, obeying
\begin{equation}
\{\Gamma_i,\Gamma_j\}=2\delta_{ij} \;,
\end{equation}
thus coordinatizing the Clifford algebra ${\cal C}\ell_N$,
that we shall denote by ${\cal C}_N$ for brevity.
The dependence of the master fields on $(y^\a,z^\a)$ is
treated using symbol calculus, whereby they belong
to classes of functions (or distributions) on ${\cal Y}_2\times
{\cal Z}_2$ that can be composed using two associative products:
the standard commutative product rule, denoted by juxtaposition, 
and an additional noncommutative product rule, denoted by a $\star$.
In what follows, we shall use the normal ordered basis in which 
the star product rule is defined formally by
\begin{equation}
(f\star g)(y,z):=\int_{\Real^4}\frac{d^2ud^2v}{(2\pi)^2}
e^{iv^\a u_\a} f(y+u,z+u)\, g(y+v,z-v)\;,
\label{hsstar}
\end{equation}
whereas a more rigorous definition requires a set of fusion 
rules (see below).
In particular, the above composition rule rigorously
defines the associative Weyl algebra ${\rm Aq}(4)$.
This algebra consists of arbitrary polynomials in $y^\a$ and $z^\a$, modulo
\begin{equation}
y_\a\star y_\b=y_\a y_\b+ i\e_{\a\b}\ ,\qquad y_\a\star z_\b=y_\a z_\b- i\e_{\a\b}\ ,\ee
\be z_\a\star y_\b=z_\a y_\b+ i\e_{\a\b}\ ,\qquad z_\a\star z_\b=z_\a z_\b- i\e_{\a\b}\ ,\ee
whose symmetric and anti-symmetric parts, respectively,
define the normal order and the (ordering independent) 
commutation rules, \emph{viz.}\footnote{The doublet variables
$y^\a$ and $z^\a$ form Majorana spinors once the 
equations are cast into a manifestly Lorentz covariant form.}
\be [y^\a,y^\b]_\star=-[z^\a,z^\b]_\star=2i\e^{\a\b}\ ,\quad [y^\a,z^\b]_\star=0\ .\ee
The basis one-forms $(dx^\mu,dz^\a)$ obey
\begin{equation}
[dx^\mu,f]_\star=0=[dz^\a,f]_\star\;,
\end{equation}
where the graded star commutator\footnote{We will explicitly 
use commutators and anticommutators, in Section \ref{sec:BulkAction}.} 
of differential forms is given by 
\begin{equation}
[f,g]_\star=f\star g-(-1)^{{\rm deg}(f){\rm deg}(g)}g\star f\;,
\end{equation}
with deg denoting the total form degree on ${\cal M}_3\times{\cal Z}_2\,$.
To describe bosonic models, we impose 
\begin{equation} \pi(A)=A\;,\quad \pi(B)=B\; \end{equation}
where $\pi$ is the automorphism of the differential star product algebra defined by
\begin{equation}
\pi(x^\mu,dx^\mu,z^\a, dz^\a,y^\a,\Gamma_i)=(x^\mu,dx^\mu,-z^\a, -dz^\a,-y^\a,\Gamma_i)\;.
\end{equation}
The hermitian conjugation is defined by 
\begin{equation}
(f\star g)^\dagger = 
(-1)^{\rm{deg}(f)\rm{deg}(g)} g^\dagger \star f^\dagger\ ,\qquad 
(z_\a,dz^\a;y_\a,\Gamma_i)^\dagger=(-z_\a,-dz^\a;y_\a,\Gamma_i)\ .\ee
and the reality conditions on the master fields read
\be
A^\dagger = -A\ ,\qquad B^\dagger = B\ .
\ee
Defining
\be F= dA+ A\star A\ ,\quad D B= dB+ A\star B-B\star A\ ,\quad
d=dx^\mu\de_\mu+dz^\a\frac{\de}{\de z^\a}\ ,\ee
where the differential obeys
\begin{eqnarray}
d( f\star g)~=~(d f)\star g+(-1)^{\rm deg(f)} 
f\star d g\ ,
\quad 
(df)^\dagger=d(f^\dagger)\;,
\end{eqnarray}  
the PV field equations can be written as
\begin{equation}\label{geomPV}
\begin{split}
&F+ B\star J=0\ ,\qquad DB=0\;,
\end{split}
\end{equation}
where 
\be J:=-\tfrac{i}{4}\,dz^\a dz_\a\,\kappa\;\qquad 
\kappa:=e^{iy^\a z_\a}\;.
\ee
The element $J$ is closed and central in the space of $\pi$-invariant forms, \emph{viz.}
\be dJ=0\ ,\qquad J\star f=\pi(f)\star J\ ,\ee
as can be seen from the fact that $\kappa$, 
which is referred to as the inner Klein operator, 
obeys
\be \k\star f(x,dx,z,dz,y,\Gamma_i)\star\k=f(x,dx,-z,dz,-y,
\Gamma_i)\ .
\end{equation}
It follows that \eqref{geomPV} defines a universally Cartan 
integrable system (\emph{i.e.} a set of generalized curvature
constraints compatible with $d^2\equiv0$ in any dimension).
The Cartan gauge transformations take the form
\begin{equation}
\delta_\e A= d\e+[A,\e]_\star\;,\quad\delta_\e B=[B,\e]_\star\;.
\end{equation}
%
\subsection{Lorentz covariance} 

Introducing 
\be S_\a:=z_\a-2i\,V_\a\ ,\qquad d_X:=dx^{\mu} \partial_\mu\ ,\ee
the equations can be rewritten as 
\begin{equation}\label{oscPV}
\begin{split}
&d_X W+W\star W=0\;,\quad d_XB+[W,B]_\star=0\;,\quad d_X S_\a+[W,S_\a]_\star=0\;,\\[2mm]
&[S_\a,B]_\star=0\;,\quad[S_\a,S_\b]_\star=-2i\e_{\a\b}\,\big(1-B\star\k\big)\;.
\end{split}
\end{equation}
In view of $\{S_\a,\k\}_\star=0$, which follows from the
bosonic projection, the above equations define a deformed 
oscillator algebra, fibered over ${\cal M}_3$, for which 
$B$ plays the role of deformation parameter. 
The equations can be cast into manifestly Lorentz
covariant form \cite{Vasiliev:1999ba,Sezgin:2002ru} by introducing a bona fide Lorentz  connection $\omega^{\a\b}=dx^\mu \omega_\mu^{\a\b}$ 
on ${\cal M}_3$ and defining 
\be {\cal W}=W-\frac{1}{4i}\omega^{\a\b} M_{\a\b}\ ,\quad   M_{\a\b}=
y_{(\a}\star y_{\b)}-z_{(\a}\star z_{\b)}+S_{(\a}\star S_{\b)}\ .\ee
in terms of which the master field equations on ${\cal M}_3$ take the form
\be \nabla {\cal W}+{\cal W}\star {\cal W} +\frac{1}{4i}r^{\a\b} M_{\a\b}=0\ ,\quad
\nabla  B+[{\cal W},B]_\star=0\;,\quad \nabla  S_\a+[{\cal W},S_\a]_\star=0\ ,\ee
where 
\be \nabla {\cal W}=d_X {\cal W}+ [\omega,{\cal W}]_\star\ ,\qquad
\nabla B=d_X B+ [\omega,B]_\star\ ,\ee
\be 
\nabla S_\a =d_X S_\a-\omega_\a{}^{\b} S_\b+ [\omega,S_\a]_\star\ .\ee
and
\be \omega=\frac{1}{4i} dx^\mu \omega^{\a\b}_\mu(y_\a\star y_\b-z_\a\star z_\b)\ ,\qquad
r^{\a\b}=d_X \omega^{\a\b}-\omega^{\a\c}\omega_{\c}{}^{\b}\ ,\ee
which are related by $r=\frac{1}{4i} r^{\a\b}(y_\a\star y_\b-z_\a\star z_\b)=
d_X\omega+\omega\star \omega$.
The deformed Lorentz generators obey the algebra
\be [M_{\a\b},M_{\c\d}]_\star =4i\e_{(\b|(\c} M_{\d)|\a)} -\delta_{\a\b} M_{\c\d}+
\delta_{\c\d} M_{\a\b}\ ,\ee
where the induced transformations
\be \delta_{\a\b} M_{\c\d}= 4i\e_{(\b|(\c} M_{\d)|\a)}- 
[y_{(\a}\star y_{\b)}-z_{(\a}\star z_{\b)}, M_{\c\d}]_\star\ee
act on the component fields of $M_{\c\d}$.
The above commutation rules are an example of a more general
construction wherein a Lie algebra $L$ acts on a space $M$
via Lie derivatives and 
\be T:L\times M\rightarrow {\cal A}\ ,\qquad T:(X,p)\mapsto T_X(p)\ ,\ee
is a representation of $L$ in an associative algebra with product $\star$
obeying
\be [T_X,T_Y]_\star = T_{[X,Y]}-{\cal L}_X T_Y+{\cal L}_Y T_X\ ,\ee
which can be seen to obey the Jacobi identity using 
$[{\cal L}_X,{\cal L}_X]={\cal L}_{[X,Y]}$ and the Leibniz' rule
${\cal L}_X (T_Y \star T_Z)= ({\cal L}_X T_Y)\star T_Z+T_Y\star ({\cal L}_X T_Z)$.

\subsection{Original PV model and its truncations} 

By taking $N=4$ and identifying 
\be \left(k\right)_{\rm PV}=\Gamma\;,\quad
\left(\nu\right)_{\rm PV}=-\nu\;,\quad
\left(\rho\right)_{\rm PV}=\C_1\;,\quad 
\left(y_\a\right)_{\rm PV}=\C_1\,y_\a\;,\quad \left(z_\a\right)_{\rm PV}=\C_1\,z_\a\;,\label{map1}\ee
\be \left(\psi_1\right)_{\rm PV}=i\Gamma_{23}\ ,\qquad \left(\psi_2\right)_{\rm PV}=i\Gamma_{24}\ .
\label{map2}\ee
we recover the original PV system, in which $\psi_1$ is
used to define the AdS$_3$ translation operators. 
By imposing the following conditions on the master fields, 
conditions that will be justified later on from the existence
of an action principle, 
\be [\Gamma, A]=0\ ,\qquad [\Gamma, B]=0\ ,\ee
\emph{i.e.} by taking them to be valued in the subalgebra  
\be {{\cal C}}^+_4= \bigoplus_{\s=\pm}
\Pi^\s_\Gamma\ {{\cal C}}_4\ \Pi^\s_\Gamma\ ,\qquad 
\Pi^\s_\Gamma=\frac12(1+\sigma \Gamma)\ ,\qquad \C=\C_{1234}\ ,\ee
of ${{\cal C}}_4$, we obtain the $\rho\,$-projected 
PV system in which the master fields $(W,B)_{\rm PV}$ 
are $\rho\,$-independent and $(S_\a)_{\rm PV}$ depend 
linearly on $\rho$. 
The $B$ field consists of eight real zero-form master fields.
Four of these describe real propagating scalar
fields in AdS$_3\,$.
The remaining four provide topological
deformation parameters. 
The following truncation 
\be (A,B)=\Pi^+_\Gamma (A,B)\ ,\ee
yields a model containing two real 
propagating scalars and two topological master fields.
Truncating one last time by imposing 
\be \tau(A,B)=(-A,B)\ ,\ee
using the anti-automorphism defined by 
\be \tau(f\star g) = (-1)^{\rm{deg}(f)\rm{deg}(g)} \tau(g) \star \tau(f)\ ,\ee
\be 
\tau(z^\a,dz^\a;y^\a,\Gamma_i)
=(-iz^\a,-idz^\a;iy^\a,\e_{(i)} \Gamma_i)\ ,\qquad \e_{(i)}=(+,+,-,-)\ ,\ee
yields a model with one propagating scalar and one topological
master field.
This model is identical to the original PV model based on
the algebra $ho_{01}^+(1,0|4)$\footnote{In the notation
of Prokushkin and Vasiliev, the algebra $ho_{01}^+(1,0|4)$ 
arises in Section 9 of \cite{Prokushkin:1998bq}
by taking $n=1$, $m=0$, $\alpha=0$ and 
$\beta=1$ in Eqs. (9.6)--(9.11) followed by 
a ${\cal P}^+$-projection (which corresponds 
to our $\Pi^+_\Gamma$-projection) and 
the projection using the anti-automorphism $\sigma$
in Eq. (4.21).}.
In all of the above models, the component along $\Gamma$ 
plays the role of a real mass parameter, denoted by $\nu$;
see Appendix A.

\subsection{Associative bundle} 

The master fields
equations define an associative algebra bundle $\widehat {\cal A}$ 
over ${\cal M}_3$ \cite{Sezgin:2011hq,Vasiliev:2015wma}. 
The fiber algebra $\widehat {\cal A}|_{p}$ at a generic point 
$p\in{\cal M}_3$ is related by a similarity transformation 
to that at a reference point $p_0 \in {\cal M}_3$.
To describe the latter, it is convenient to separate the variables 
in $\Omega_{[0]}({\cal Y}_2) \otimes \Omega({\cal Z}_2)$,
as \eq{hsstar} reduces to separate Weyl order on 
$\Omega_{[0]}({\cal Y}_2)$ and $\Omega({\cal Z}_2)$.
One can then use the factorization formula 
\be \label{factorkappa}
\kappa=\kappa_z \star \kappa_y\ ,\qquad \kappa_y
:=2\pi\,\delta^2(y^\a)\ ,\qquad
\kappa_z:=2\pi\delta^2(z^\a)\ ,
\ee
to solve the deformed oscillator algebra 
at $p_0$ formally in terms of auxiliary integrals 
facilitated by analytical continuation 
methods in $\Omega({\cal Z}_2)$
\cite{Iazeolla:2011cb}.
Thus,
\be 
\widehat {\cal A}|_{p}\cong 
\widehat{\cal A}\rvert_{p_0}=
\Omega({\cal Z}_2)
\otimes {\cal A}\otimes {{\cal C}}_N\ ,\qquad
{\cal A}=\bigoplus_{\Sigma} {\rm Aq}(2)[\Sigma]\ ,
\label{widehatcalAfiber}
\ee
where ${\rm Aq}(2)[\Sigma]$ are vector spaces
of symbols corresponding to a set of boundary 
conditions on ${\cal M}_3\times {\cal Z}_2$ 
\cite{Iazeolla:2008ix,Iazeolla:2011cb}; for
examples, see Section 4.
Harmonic expansions, spectrum analysis  
and exact solutions show that the
associative bundle contains nonpolynomial 
sectors obtainable from reference elements 
\cite{Iazeolla:2008ix,Iazeolla:2011cb,Boulanger:2015kfa}
\be T_\Sigma \in {\rm Aq}(2)[\Sigma]\ ,\ee
by the left and right action of the Weyl algebra ${\rm Aq}(2)$.
We write
\be {\rm Aq}(2)[\Sigma]={\rm Aq}(2)[T_\Sigma;\l,\rho]\ ,\ee
indicating the properties of ${\rm Aq}(2)[\Sigma]$ 
as a left ($\lambda$) and right ($\rho$) module 
of ${\rm Aq}(2)$.
The associative structure of ${\cal A}$ requires a 
fusion rule 
\be 
{\rm Aq}(2)[\Sigma] \star {\rm Aq}(2)[\Sigma']=\bigoplus_{\Sigma''}
{\cal N}_{\Sigma\Sigma'}{}^{\Sigma''}{\rm Aq}(2)[\Sigma'']\ ,\qquad
 {\cal N}_{\Sigma\Sigma'}{}^{\Sigma''}\in \{0,1\}\ ,
 \label{fusion}
 \ee
such that if ${\cal N}_{\Sigma\Sigma'}{}^{\Sigma''}=1$ then
the left-hand side is to be computed using \eq{hsstar} 
with $z^\a=0$ and expanded into the basis of 
${\rm Aq}(2)[\Sigma'']$ such that all nontrivial
products are finite and the resulting multiplication
table is associative.
For example, massless particles and various types of
algebraically special exact solution spaces arise 
within Gaussian sectors.
The Weyl algebra
${\rm Aq}(2)\equiv {\rm Aq}(2)[\mathbf 1]$, with 
reference state being the identity operator, 
is also included (as the sector corresponding to 
twistor space plane waves), typically with\footnote{An exception
is fractional spin gravity \cite{Boulanger:2013naa} whose
fractional spin sector $\Psi$ and 
Lorentz singlet sector $U$ have ${\cal N}_{\Psi1}{}^{\Psi}=1$
and ${\cal N}_{1U}{}^{U}={\cal N}_{U 1}{}^{U}=0$.}
${\cal N}_{1\Sigma}{}^{\Sigma}={\cal N}_{\Sigma 1}{}^{\Sigma}=1$.
%

%
\section{Covariant Hamiltonian action}
\label{sec:BulkAction}
%

In this section we begin by discussing some generalities on covariant Hamiltonian actions on ${\cal X}_4\times {\cal Z}_2$. We then determine the constraints on the Hamiltonian such that it leads to a master action in which the master field content, including the Lagrange multipliers, are extended to consist of sum of even and odd forms of appropriate degree, and central elements. This action yields a generalized version of the PV field equations.

\subsection{Generalities} 

In order to formulate the theory 
within the AKSZ framework \cite{Alexandrov:1995kv} using its
adaptation to noncommutative higher spin geometries proposed
in \cite{Boulanger:2012bj}, we assume a formulation of the PV 
system that treats ${\cal Z}_2$ as being closed and introduce 
an open six-manifold ${\cal M}_6$ with boundary
\be \partial{\cal M}_6={\cal X}_3\times{\cal Z}_2\ ,\ee
where ${\cal X}_3$ is a closed manifold containing ${\cal M}_3$ 
as an open submanifold. 
On ${\cal M}_6$, we introduce a two-fold duality extended
\cite{Vasiliev:2007yc,Boulanger:2011dd,Vasiliev:2015mka}
\footnote{Starting from a universally 
Cartan integrable system and replacing each $p$-form by a sum of forms 
of degrees $p$, $p+2$, $\dots$, $p+2N$, and each structure constant 
by a function of off-shell closed and central terms,
\emph{i.e.} elements in the de Rham cohomology valued in 
the center of the fiber algebra, with a decomposition
into degrees $0$, $2$, $\dots$, $2N$, yields a new universally Cartan 
integrable system, referred to as the $N$-fold
duality extension of the original system.
More generally, one may consider on-shell duality
extensions by including on-shell closed complex-valued 
functionals into the extension of the structure constants 
\cite{Sezgin:2012ag,Vasiliev:2015mka}.} set of differential forms given by
\begin{eqnarray}
A &=& A_{[1]}+A_{[3]}+A_{[5]}\;,\qquad 
B = B_{[0]}+B_{[2]}+B_{[4]}\;,
\\
T &=& T_{[4]}+T_{[2]}+T_{[0]}\;,\qquad S = S_{[5]}+S_{[3]}+S_{[1]}
\;,
\end{eqnarray}
valued in ${\cal A}\otimes {\cal C}_4$ and where the subscript denotes the form degree.
We let $\{ J^{I}\}$ denote the generators of the ring 
of off-shell closed and central terms, \emph{i.e.}
elements in the de Rham cohomology of ${\cal M}_6$
valued in the center of ${\cal A}\otimes {\cal C}_4$, which hence obey
\begin{eqnarray}
{d} J^I~=~0\ ,\quad \left[ J^I, f\right]_\star~=~0\ ,
\end{eqnarray} 
(off-shell) for any differential form $f$ on ${\cal M}_6$
valued in ${\cal A}\otimes {\cal C}_4$.
Following the approach of \cite{Boulanger:2011dd}, we consider actions
of the form
\begin{eqnarray}
S_{\rm H}~&=&~\int_{{\cal M}_6} {\rm Tr}_{{\cal A}\otimes {\cal C}_4}
\left[  S\star D B+ T\star  F+{\cal V}( S, T; B; J^I)\right]
\label{action1}
\\
&=&\int_{{\cal M}_6}{\rm Tr}_{{\cal A}\otimes {\cal C}_4}
\left[  S\star {\rm d} B+ T\star  {\rm d}A-{\cal H}( S, T;A, B; J^I)\right]
\label{action2}
\end{eqnarray}
where ${\rm Tr}_{{\cal A}\otimes {\cal C}_4}$ denotes
a cyclic trace operation 
on ${\cal A}\otimes {\cal C}_4\,$.
We assume a structure group gauged by $A$ and that 
$S$, $T$ and $B$ belong to sections, and \eq{action2} 
makes explicit the covariant Hamiltonian form, with
\begin{equation}
\begin{split}
&{\cal H}(S,T;A,B;J^I)=-S\star[A,B]_\star-T\star A\star A-{\cal V}(S,T;B;J^I)\;.
\end{split}
\end{equation}
Thus, the coordinate and momentum master fields, defined by
\be  
(X^\alpha;P_\alpha):=(A,B;T,S)\ ,
\ee
lie in subspaces of ${\cal A}$ that are dually paired 
using ${\rm Tr}_{\cal A}$, which leads to distinct 
models depending on whether these subspaces are isomorphic 
or not. 
In the reductions that follow, we shall consider the
first type of models, while a model with coordinates
and momenta in non-isomorphic spaces is treated in 
Appendix B.
Moreover, for definiteness, we shall assume that
\be 
{\cal M}_6={\cal X}_4\times {\cal Z}_2\ ,\ee
and the associative bundle $\widehat{\cal A}$ defined
in \eq{widehatcalAfiber} is chosen such that
\be 
\check {\cal L}=\oint_{{\cal Z}_2} {\rm Tr}_{{\cal A}\otimes {\cal C}_4}
\left[  S\star D B+ T\star  F+{\cal V}( S, T; B; J^I)\right]\ ,
\ee
is finite (and globally defined on ${\cal X}_4$).
The action can then be written as
\be S_{\rm H} = \int_{{\cal X}_4} 
\check{\cal L}\ .
\ee
We shall furthermore assume that
\be \int_{{\cal M}_6} {\rm Tr}_{{\cal A}\otimes {\cal C}_4} {d} f=
\oint_{\partial {\cal M}_6} {\rm Tr}_{{\cal A}\otimes {\cal C}_4}
f\ ,\ee
and  
\be
\int_{{\cal M}_6} {\rm Tr}_{{\cal A}\otimes {\cal C}_4} f\star g = (-1)^{{\rm deg}(f){\rm deg}(g)}
\int_{{\cal M}_6} {\rm Tr}_{{\cal A}\otimes {\cal C}_4} g\star f\ ,\ee
\be 
\oint_{\partial {\cal M}_6} {\rm Tr}_{{\cal A}\otimes {\cal C}_4}
f\star g~=~\oint_{\partial{\cal M}_6} {\rm Tr}_{{\cal A}\otimes {\cal C}_4} 
g\star f \ ,
\ee
from which it follows that ${\cal H}$ is a graded 
cyclic $\star$-function.

\subsection{Constraints on ${\cal H}$}

The Hamiltonian is constrained by gauge invariance, 
or equivalently, by universal on-shell Cartan integrability\footnote{
Covariant Hamiltonian actions are gauge invariant iff their 
equations of motion form universally Cartan integrable systems.}.
In addition, it is constrained by the requirement
that the equations of motion on ${\cal M}_6$ 
reduce to a desired set of equations of motion 
on $\partial{\cal M}_6$ upon assuming 
natural boundary conditions.
To examine the above, we let 
\be Z^i\equiv (X^\alpha;P_\alpha)\ ,\ee
and consider the total variation
\begin{eqnarray}
\delta S_{\rm H}&\equiv &\int_{{\cal M}_6} {\rm Tr}_{{\cal A}\otimes {\cal C}_4} 
\delta Z^i \star {\cal R}^j\, {\Omega}_{ij}
 +(-)^{{\rm deg}(P_\alpha)}\oint_{\partial 
 {\cal M}_6} {\rm Tr}_{{\cal A}\otimes {\cal C}_4} P_\alpha\star\delta X^\alpha\ ,
 \label{variationaction}
\end{eqnarray}
where $\Omega_{ij}$ is a graded anti-symmetric constant 
matrix\footnote{Adopting the conventions of \cite{Boulanger:2012bj},
we take $\Omega^{ik}\Omega_{kj}=-\delta^i_j\,$.} and the 
Cartan curvatures are given by
\begin{eqnarray}
{\cal R}^i~:=~
{d}Z^i+{\cal Q}^i(Z)~\approx~0\ ,\quad {\cal Q}^i~:=~\Omega^{ij} \del_j {\cal H}\ ,
\end{eqnarray} 
where $\del_i$ denotes the graded cyclic derivative defined by
\be \delta \int_{{\cal M}_6} {\rm Tr}_{{\cal A}\otimes {\cal C}^+_4}
{\cal U}~=~\int_{{\cal M}_6} {\rm Tr}_{{\cal A}\otimes {\cal C}^+_4} \delta Z^i\star \del_i {\cal U}\ ,\ee
for any graded cyclic $\star$-function ${\cal U}$. 
We find 
\bea 
&& {\cal R}^A ~=~ F+\del_T{\cal V}  \ ,
\qquad {\cal R}^B ~=~ DB+\del_S{\cal V} \ ,
\\[5pt]
&& {\cal R}^S~=~DS+\del_B{\cal V} \ ,
\qquad {\cal R}^T~=~DT-[B,S]_\star \ .
\eea
Requiring $A$ and $B$ to be free to fluctuate on 
$\partial {\cal M}_6\,$, the variational principle implies
\begin{equation}
\left.P_\alpha\right\vert_{\partial {\cal M}_6} = 0 ~.  
\end{equation}
The Cartan integrability requires 
\be 
\overrightarrow{\cal Q}\star {\cal Q}^i 
~\equiv~0 \ ,\label{quadratic}
\ee
using a notation in which $\star$-vector fields  
$\overrightarrow V\equiv V^i \overrightarrow\del_i$ act on 
$\star$-functions as follows:
\footnote{If ${\cal U}_{\rm symm}$ is a totally symmetric $\star$-function, then
$\del_i{\cal U}_{\rm symm}~=~\overrightarrow \del_i{\cal U}_{\rm symm}$.
}
\be
\overrightarrow V \star ({\cal U}_1\star {\cal U}_2)
= (\overrightarrow V \star {\cal U}_1)\star {\cal U}_2
+(-1)^{{\rm deg}(\overrightarrow V){\rm deg}({\cal U}_1)}\;
{\cal U}_1\star (\overrightarrow V \star {\cal U}_2)\ ,\qquad \overrightarrow V Z^i = V^i\ .
\ee
Moreover, imposing
\be \left.\del_i {\cal V}\right|_{ P_\a=0}~=~(0,0;{\cal F},0) \ ,\label{linear}
\ee
the set of boundary equations is a 
two-fold duality extension of the PV system, \emph{viz.}  
\begin{equation}
F+{\cal F}( B; J^I)=0\ ,\qquad  D B=0\ ,\label{eom}
\end{equation} 
where  
\begin{eqnarray}
{\cal F}( B; J^I)~:=~\sum_{n\geqslant 0} {\cal F}_n(J^I)\star B^{\star n}\ ,\quad 
{\cal F}_{n}(J^I)~=~\sum_{k\geqslant 0} {\cal F}_{n,I_1\dots I_k} J^{I_1}\star\cdots\star  J^{I_k}\ ,
\end{eqnarray} 
for a set of complex constants ${\cal F}_{n,I_1\dots I_k}$.
%
\subsection{The master action}

In order to obtain a model that admits consistent truncations
to three-dimensional CS higher spin gravities, we need to 
assume that ${\cal V}$ contains a term that is quadratic in $T$.
The simplest possible such action is given by
\be
\boxed{
S_{\rm H} = \int_{{\cal M}_6} {\rm Tr}_{{\cal A}\otimes {\cal C}_4}
\left[ S\star DB + T\star\left[F+g+h\star ( B - \tfrac12\mu\star T) \right] 
+\mu \star B\star S\star S \right]}
\label{ma}
\ee
where 
\be 
g=g(J^I)\ ,\qquad h=h(J^I)\ ,\qquad \mu=\mu(J^I)
\ee
are even closed and central elements on ${\cal M}_6$
in degrees
\be {\rm deg}(g,h,\mu)=(2\ \mbox{mod $2$},2\ \mbox{mod $2$},0\ \mbox{mod $2$})\ .\ee
The reality conditions are given by
\be (A,B;T,S;g,h,\mu)^\dagger=(-A,B;-T,S;-g,-h,-\mu)\ ,\ee
The total variation 
\begin{eqnarray}
\delta S_{\rm H} &=& \int_{{\cal M}_6} 
{\rm Tr}_{{\cal A}\otimes{\cal C}_4}\Big( 
\delta T \star{\cal R}^A +\delta S\star {\cal R}^B +\delta A\star {\cal R}^T +\delta B \star{\cal R}^S\Big)
\nonumber\\
&& +\;\oint_{\partial{\cal M}_6}{\rm Tr}_{{\cal A}\otimes{\cal C}_4}(T\star\delta A - S\star \delta B) \;,
\end{eqnarray}
where the Cartan curvatures read

\medskip

\be
\begin{aligned}
{\cal R}^A &=F + g+ h\star ( B - \mu\star T)\approx 0 
\w2
{\cal R}^B &= DB + \mu \star [S,B]_\star\approx 0
\w2 
{\cal R}^T &= DT + [S,B]_\star\approx 0 
\w2
{\cal R}^S &= DS +h\star T+ \mu \star S\star S\approx0
\label{e4}
\end{aligned}
\ee
\medskip
The generalized Bianchi identities are
\begin{eqnarray}
D{\cal R}^A &\equiv & h\star ( {\cal R}^B  - \mu\star {\cal R}^T)  \;,
\\
D{\cal R}^B &\equiv & [ ( {\cal R}^A + \mu\star{\cal R}^S ) , B]_\star - \mu\star\{ {\cal R}^B , S \}_\star\;,
\\ 
D{\cal R}^T &\equiv & [ {\cal R}^A , T]_\star + [  {\cal R}^S , B]_\star  - \{ {\cal R}^B , S \}_\star\;,
\\ 
D{\cal R}^S &\equiv & [  {\cal R}^A , S ]_\star + \mu\star [ {\cal R}^S , S ]_\star + h\star{\cal R}^T \;.
\end{eqnarray}
The gauge transformations 
\begin{eqnarray}
\delta_{\e,\eta} A &=& D\epsilon^A - h\star(\epsilon^B -\mu\star \eta^T)\;,
\\
\delta_{\e,\eta} B &=& D\epsilon^B - [\epsilon^A , B]_\star - \mu\star[\eta^S , B]_\star + \mu \star
\{ S,\epsilon^B\}_\star \;,
\\
\delta_{\e,\eta} T &=& D\eta^T - [\epsilon^A , T]_\star - [\eta^S , B]_\star+ \{ S,\epsilon^B\}_\star  \;,
\\
\delta_{\e,\eta} S &=& D\eta^S - [\epsilon^A , S]_\star - \mu\star [\eta^S , S]_\star -  h\star \eta^T\;,
\end{eqnarray}
which transform the Cartan curvatures into each other, induce 
\begin{eqnarray}
\delta_{\e,\eta}S_{\rm H} &=& \int_{{\cal M}_6} {\rm Tr}_{{\cal A}\otimes {\cal C}_4}
\Big( \eta^T\star [F+ g+h\star B] + \eta^S \star DB \Big)\;.
\end{eqnarray}
We take $(\e^B;\eta^T,\eta^S)$ to belong 
to sections of the structure group and impose\footnote{
Following the AKSZ approach, the Batalin--Vilkovisky classical
master equation requires that the ghosts corresponding
to $(\eta^T,\eta^S)$ vanish at $\partial{\cal M}_6$ off-shell.}
\be (\eta^T,\eta^S)\rvert_{\partial{\cal M}_6}=0\ .\ee
We have also assumed that $(A,B)$ fluctuate on $\partial{\cal M}_6$, which implies\footnote{
Following the AKSZ approach, the Batalin--Vilkovisky classical
master equation requires that \eq{bc} holds off-shell.}
\be T\rvert_{\partial{\cal M}_6}\approx 0\approx S\rvert_{\partial{\cal M}_6}\ .\label{bc}\ee
The resulting boundary equations of motion 
\begin{equation}
F+g+ h\star B\approx0\;,\quad DB\approx0\;
\label{ev}
\end{equation}
thus provide a duality extended version of the 
Prokushkin--Vasiliev equations, that is free from any interaction 
ambiguity, following a variational principle.

In the action \eq{ma}, the relative coefficient of the
$BSS$ and $TT$ terms is fixed uniquely by Cartan integrability.
The action is invariant under $(B,S;\mu,h)\rightarrow (\lambda\star B,\lambda^{-1}\star S; \lambda\star \mu,
\lambda^{-1}\star h)$ for closed and central elements $\lambda=\lambda(J^i)$ of degree
$0$ mod $2$ that are real and invertible.
The canonical transformation  
$(A,B)\rightarrow(A -\tfrac12 \mu\star S,
B+\tfrac12\mu\star T)$ leads to replacement of
$\int_{{\cal M}_6} 
{\rm Tr}_{{\cal A}\otimes {\cal C}_4}
\left[ - \tfrac12\mu\star T\star T
+\mu \star B\star S\star S \right]$ in \eq{ma} by  
 $\tfrac14 \int_{{\cal M}_6}
{\rm Tr}_{{\cal A}\otimes {\cal C}_4}
\mu\star T\star S\star S-\tfrac12 
\oint_{\partial{\cal M}_6}
{\rm Tr}_{{\cal A}\otimes {\cal C}_4}\mu\star T\star S$.
However, as we shall see, the form of the
Hamiltonian action for the PV system that
lends itself most straightforwardly 
to consistent truncations of the $B$ field 
is given by \eq{ma}.

\section{Consistent truncations}\label{Sec:Blencowe}

In this section we perform consistent truncations of the 
covariant Hamiltonian master action in six dimensions down
to various models on ${\cal X}_4$ and ${\cal Z}_2$.
The truncations consist of integrating out the fluctuations in 
$B$ around its vacuum expectation value $\nu\Gamma$
followed by reductions on ${\cal Z}_2$ and ${\cal X}_4$.
On ${\cal X}_4$, we reach $BF$-like models with Lagrangian forms 
containing Blencowe's action for $\nu=0$ and a $\nu$-deformed 
version thereof that we present in Appendix B.
For $\nu=0$, the reduction to ${\cal Z}_2$ yields the 
Prokushkin--Segal--Vasiliev (PSV) action.

A consistent truncation a system with action 
$S[\varphi]$ and equations of motion $E(\varphi)= 0$ amounts 
to an Ansatz $\varphi=\varphi(\varphi^{\prime})$ off-shell such that 
$E(\varphi(\varphi'))=0$ are equivalent to a set of equations 
$E'(\varphi')=0$ that i) are integrable without any algebraic 
constraints on $\varphi'$; and ii) follow by applying the
variational principle to the reduced action  
$S_{\rm red}[\varphi']:=S[\varphi(\varphi')]$.
%

\subsection{Reduction to BF-like extension of Blencowe's action}

Starting from the equations of motion \eq{e4} and setting
$B=0$ yields 
\be F + g- h\star \mu\star T=0 \ ,\qquad  DT = 0\ ,\label{sub}\ee
and
\be 
DS +h\star T+ \mu \star S\star S=0\ , 
\label{ds}
\ee
which together form a Cartan integrable system containing 
\eq{sub} as a subsystem, \emph{i.e.}
the free differential algebra generated by $(A,T,S)$
contains a subalgebra generated by $(A,T)$.
Assuming $\partial {\cal M}_6$ to consist 
of a single component, it follows from 
$S\vert_{\partial {\cal M}_6} =0$ that $S$
can be reconstructed from $(A,T)$ on-shell
\footnote{
Since $T\vert_{\partial {\cal M}_6} =0$
on-shell as well it follows that both $S$
and $T$ can be taken to vanish on ${\cal M}_6$
on-shell.} from \eq{ds}.
Therefore, the system \eq{sub} is a consistent 
truncation of the original system \eq{e4} on-shell.

Rewriting the full action \eq{ma} by integrating
by parts in its $SDB$-term yields
\be 
S_{\rm H}=\int_{{\cal M}_6} {\rm Tr}_{{\cal A}\otimes {\cal C}_4}
\left[T\star (F+g-\tfrac12 h\star \mu \star T)+B\star (DS+h\star T+\mu\star S\star S)\right]\ .
\ee
It follows that $B=0$ is a saddle point of
the path integral at which $B$ and $S$
can be integrated out in a perturbative
expansion.
Schematically, modulo gauge fixing, one has
\be \int_{\langle B\rangle =0} [DB][DS] e^{\tfrac{i}\hbar S_{\rm H}}
\sim  e^{\tfrac{i}\hbar S_{\rm eff}[A,T]}\ ,\ee
where the effective action
\be S_{\rm eff}[A,T]=S_{\rm red}[A,T]+ O(\hbar)\ ,\label{Seff}\ee
consists of loop corrections (comprising 
attendant functional determinants on noncommutative 
manifolds) and 
\be S_{\rm red}=\int_{{\cal M}_6} {\rm Tr}_{{\cal A}\otimes {\cal C}_4}
T\star (F+g-\tfrac12 h\star \mu \star T)\ .\label{Sred1}\ee
The latter is a consistently reduced classical action
in the sense that it reproduces the subsystem \eq{sub}.
The reduced system, which thus consists if the 
gauge sector of the original system, is a topological 
theory with local symmetries
\begin{equation}
\delta A=D\e+\mu\star h\star\eta\;,\quad\delta T
=D\eta-[\e,T]_\star\;,
\end{equation}
and equations of motion and boundary conditions given by
\bea
&& F+g-\mu\star h\star T=0\;,\quad DT=0\;,
\\
&& T\rvert_{\partial{\cal M}_6}=0\;.
\label{tbc}
\eea
The boundary equations are thus given by 
\be (F+g)\rvert_{\partial{\cal M}_6}=0\ .\ee

To address Blencowe's theory, we truncate once 
more by reducing \eq{sub} under the assumptions 
that   
\be 
g=\check g_{[2]} - \mu_0 J \star \check g'_{[2]}\ ,\quad \mu =\mu_{[0]}\equiv \mu_0\ ,\quad
h=J\ ,
\label{reduceg}
\ee
where $\mu_0$ is an imaginary constant, and that
\be 
A = \check{W}_{[1]} - \check K_{[1]}-\mu_0 J \star \check{K}_{[1]}\ ,\label{r1}\ee\be  
T = \check{T}_{[2]}+ \check K_{[1]}\star \check K_{[1]} -
\mu_0 J\star \check{T}_{[2]}\;,
\label{r2}
\ee
where by definition 
\be 
\check{f}\in \Omega({\cal X}_4) \otimes 1_{\Omega({\cal Z}_2)}\otimes \check{\cal A}\otimes 
\check{\cal C}_4 \ ,
\label{r3}
\ee
in terms of an associative algebra $\check{\cal A}$ of 
$\pi$-projected symbols of $y^\a$ (to be specified below).
Thus 
\be d\check{f}=d_X\check f\ ,\qquad \pi(\check{f})=\check f\ ,\ee
as required for $\pi(A,T)=(A,T)$.
Defining 
\be \check F=d_X \check{W}+\check{W}\star \check{W}\ ,\quad 
\check D\check K=d_X \check{K}+[\check{W},\check{K}]_\star\ ,\quad 
\check D\check T=d_X \check{T}+[\check{W},\check{T}]_\star\ ,\label{defcheck}\ee
suppressing the subscripts indicating form degrees,
the reduction of \eq{sub} yields 
\be 
\check F+ \check  T+\check g +\check g' =0\ ,\qquad 
\check D\check  T=0\ ,\label{red1}
\ee
\be 
D \check K-\check K\star \check K+\check T+
\check g'=0\ ,
\label{red2}
\ee
which is a Cartan integrable system containing \eq{red1}
as a subsystem.
From \eq{tbc} and \eq{r2}, we deduce the boundary conditions 
\be \check  T\rvert_{\partial{\cal X}_4}=0=(\check K\star \check K)\rvert_{\partial{\cal X}_4}\ ,
\label{red3}\ee
which are compatible with \eq{red2} since $[\check g',\check K]_\star=0$.
Substituting \eq{r1} and \eq{r2} into \eq{Sred1} 
and using \eq{red3} we obtain 
\be
\check S_{\rm red}[\check W,\check T]=-\mu_0 \int_{{\cal X}_4} \int_{{\cal Z}_2}
{\rm Tr}_{{\cal A}\otimes {\cal C}_4}
J\star \check {T}\left( \check{F}+\check g +\check g '+
\tfrac12 \check{T}\right)\ ,\label{checkSred}\ee
which reproduces \eq{red1} on-shell, implying that
truncation \eq{r1}--\eq{r3} is indeed consistent.

There are two independent embeddings of Blencowe's model into the above master action.
They can be obtained by choosing the fiber algebras
\be 
m=0:\quad \check{\cal A}\otimes \check {\cal C}_4=\left({\rm Aq}^+(2)\oplus 
({\rm Aq}^+(2)\star \kappa_y)\right)\otimes {\cal C}_4\ ,\ee
\be m=1:\quad \check{\cal A}\otimes \check {\cal C}_4=\left({\rm Aq}^+(2)\oplus 
({\rm Aq}^+(2)\star \kappa_y)\right)\otimes {\cal C}^+_4\ ,\ee
and equipping ${\cal A}\otimes {\cal C}_4$ with trace operations 
as follows\footnote{The (graded cyclic) supertrace obeys ${\rm STr}_{{\rm Aq}(2)} f\star g= 
{\rm STr}_{{\rm Aq}(2)}  g\star \pi(f)$ and ${\rm STr}_{{\rm Aq}(2)}f=f(0)$
provided that $f(y)$ is the symbol of $f$ defined in Weyl order.}:
\be 
{\rm Tr}^m_{{\cal A}\otimes {\cal C}_4} (f_0+f_1 \star \kappa_y) :=
\int_{{\cal Y}_2}\frac{d^2y}{2\pi} {\rm Tr}_{{\cal C}_4}(\Gamma)^m \kappa_y\star f_{m}\equiv 
{\rm STr}_{{\rm Aq}(2)} {\rm Tr}_{{\cal C}_4}(\Gamma)^m f_{m}\ ,\qquad m=0,1\ ,
\label{choice}
\ee
where $f_m\in {\rm Aq}^+(2)\otimes {\cal C}_4$ and
\be  
\int_{{\cal Y}_2}\frac{d^2y}{2\pi}=\int_{\Real^2}\frac{d^2y}{2\pi}\ ,\qquad
{\rm Tr}_{{\cal C}_4}
\sum_{k=0}^4
f_{i_1..i_k}\,\C^{[i_1}..\C^{i_k]}  
:=\,f_0\;\ .
\end{equation}
The factorization formula \eq{factorkappa} then yields\footnote{We use the normalizations
$dz^\a dz_\a=-2 dz^1 dz^2=-2 d^2z$ and $\int d^2y d^2z \kappa\star f(y)=4\pi^2 f(0)\,$.}
\be
\int_{{\cal Z}_2}  {\rm Tr}^m_{{\cal A}\otimes{\cal C}_4} J\star (\check f_0+ \check f_1\star \kappa_y)=
i\pi\, {\rm STr}_{{\rm Aq}(2)}  {\rm Tr}_{{\cal C}_4}(\Gamma)^m \check f_{1-m}\ .
\ee
We truncate the models further as follows: 
\bea 
m=0&:&\quad \check W=\Pi^+_{\kappa_y}\star
W_++\Pi^-_{\kappa_y}\star W_-\ ,\qquad 
\check T=\Pi^+_{\kappa_y}\star
T_++\Pi^-_{\kappa_y}\star T_- \;,
\label{Blred1}\w2
m=1&:& \quad \check W=\Pi^+_\Gamma
W_++\Pi^-_\Gamma W_-\ ,\qquad 
\check T=\Pi^+_\Gamma
T_++\Pi^-_\Gamma T_- \;,
\label{Blred2}
\eea
where $W_\pm$ and $T_\pm$ are independent of $\Gamma_i$ and $\kappa_y$, and
\be \Pi^\pm_{\kappa_y}=\frac{1\pm\kappa_y}2\ .\ee
Inserting \eq{Blred1} and \eq{Blred2} into \eq{checkSred} and using 
\be 
m=0:\quad\int_{{\cal Z}_2}  {\rm Tr}^0_{{\cal A}\otimes{\cal C}_4} J\star \Pi^\pm _{\kappa_y}f
=\pm \tfrac{i\pi}{2} \,{\rm STr}_{{\rm Aq}(2)}f\ ,
\ee
\be 
m=1:\quad \int_{{\cal Z}_2}  {\rm Tr}^1_{{\cal A}\otimes{\cal C}_4} J\star \Pi^\pm _{\Gamma}f
=\pm \tfrac{i\pi}{2} \, {\rm STr}_{{\rm Aq}(2)}\,f\ ,
\ee
for $f$ independent of $\Gamma_i$ and $\kappa_y$,
yields the following four-dimensional Hamiltonian extension
of Blencowe's action:
\be
\label{BlenH}
S_{\rm Bl}= - \tfrac{i\pi}{2} \, \mu_0 \int_{{\cal X}_4} 
{\rm STr}_{{\rm Aq}(2)} \left[
T_+(F_++\check g +\check g '+
\tfrac12 T_+ )-T_-(F_-+\check g +\check g '+
\tfrac12 T_- )\right]\ ,
\ee
which is thus reached for both $m=0$ and $m=1$.

Assuming that ${\cal X}_4={\cal X}_3\times [0,\infty[$ and that
all fields fall off at ${\cal X}_3\times\infty$, and assuming
furthermore that ${\cal X}_3$ has a simple topology such that 
\be 
\check g +\check g '=0\ ,
\ee
the elimination of the Lagrange multipliers yields
\be
\label{BlenCS}
S_{\rm Bl}=\tfrac{i\pi}2\, \mu_0  \left(S_{\rm CS}[W_+]-
S_{\rm CS}[W_-]\right)\ ,\ee
with 
\be S_{\rm CS}[W]=\oint_{{\cal X}_3}{\rm STr}_{{\rm Aq}(2)} \Big[\tfrac12 W\star 
dW+\tfrac13W\star W\star W\Big]\;,
\ee
where now $d$ denotes the exterior derivative on ${\cal X}_3$.
Equivalently, 
\begin{equation}\label{EHaction}
S_{\rm Bl}=i\mu_0 \pi\oint_{{\cal X}_3} {\rm STr}_{{\rm Aq}(2)} \Big[E
\star (d\Omega+\Omega\star \Omega)+\tfrac13\,E\star E\star E\Big]\;,\qquad W_\pm =\Omega\pm E\ ,
\end{equation}
from which we identify
\begin{equation}
\mu_0 = -\frac{4i}{\pi^2} \frac{\ell_{\rm AdS}}{G_{\rm N}} 
\end{equation}
using the conventions of \cite{Boulanger:2015uha}. 
Relaxing the assumption on $\check g +\check g '$ by taking
it to be a nontrivial element in the de Rham cohomology 
of ${\cal X}_3$, Blencowe's action is accompanied by
the extra term 
\bea S_{g}&=&2i\pi \mu_0 \int_{{\cal X}_4}{\rm STr}_{{\rm Aq}(2)}(\check g +\check g ')\star \check F
=2i\pi \mu_0 \oint_{{\cal X}_3}{\rm STr}_{{\rm Aq}(2)} (\check g +\check g ')\star \check W\nn\w2
&=&i \pi \mu_0 \oint_{{\cal X}_3}{\rm STr}_{{\rm Aq}(2)} (\check g +\check g ')\star E
\ ,\eea
which is the flux of the central gauge fields in $\check W$ through the
two-cycle dual to $\check g +\check g'$.
Thus, the modified Blencowe equations of motion take the form
\be d\Omega+\Omega\star \Omega+ E\star E=-(\check g +\check g ')\ ,\qquad  
dE+\Omega\star E+ E\star \Omega=0\ .\ee

\subsection{Reduction to PSV action}

Instead of reducing \eqref{ma} on ${\cal Z}_2$ 
one may consider a reduction on ${\cal X}_4\,$ 
under the assumption that 
\be \partial{\cal X}_4=\emptyset\ ,\ee
as well as $\partial{\cal Z}_2=\emptyset\,$. 
The absence of any boundary condition on $T$ implies
that its integration constant in form degree zero 
contains local degrees of freedom.
To exhibit the model on ${\cal Z}_2$, 
we first perform a consistent truncation 
by setting
\be 
B=\nu \Gamma\ ,
\ee
leading to the reduced action
\be 
S_{\rm red}[A,T] = 
\int_{{\cal M}_6} {\rm Tr}_{{\cal A}\otimes {\cal C}_4}
\Big[ T\star (F+g+\nu\Gamma h-\tfrac12 h\star \mu \star T)\Big] \ .
\label{Sred1b}
\ee
We then proceed by introducing a volume form 
$\check J_{[4]}$ on ${\cal X}_4$ and background 
potentials $\check W^{(0)}_{[3]}$ and $V^{(0)}_{[1]}$ on 
${\cal X}_4$ and ${\cal Z}_2$, respectively,
defined by 
\be 
d_X\check W^{(0)}_{[3]}=\check J_{[4]}\ ,\qquad 
F^{(0)}_{[2]} + \nu \Gamma J = 0 \ , 
\qquad 
F^{(0)}_{[2]}:= d_Z V^{(0)}_{[1]}+ V^{(0)}_{[1]}\star V^{(0)}_{[1]} \ . 
\ee
In particular, we take 
$\check W^{(0)}_{[3]}$ to be independent of the 
internal coordinates $y^\alpha\,$.
We next perform a further truncation by taking
\be 
h=J+i\check J_{[4]}\ ,\qquad \mu=\mu_0\ ,\qquad g=g'_{[2]}\ ,
\ee
and considering the Ansatz
\be 
A=V^{(0)}_{[1]}+V'_{[1]} -\mu_0\, 
{\check W}^{(0)}_{[3]}\star
\left( 1 + i[\alpha -\beta] J \right)\star C' 
- i \nu\Gamma\,{\check W}^{(0)}_{[3]} \ ,\ee
\be
T=i\,(1+i\a J + \b \check J_{[4]})\star C'\ ,
\ee
with $\a,\b\in\Real$ and fluctuating fields
\be f'\in 1\rvert_{{\cal X}_4}\otimes {\Omega}({\cal Z}_2)\otimes
{\cal A}' \otimes {\cal C}^+_4\ ,\qquad 
\pi(f')=f'\ ,\qquad df' = d_Z f'\ ,\ee
where ${\Omega}({\cal Z}_2)\otimes{\cal A}'$ 
consists of an algebra of $\pi$-invariant 
master fields.
Defining 
\be F'=d_Z V'+ \{ V^{(0)}, V'{\}}_\star+V'\star V'\ ,\qquad D'C'=
d_Z C'+ [ V^{(0)}+V', C']_\star\ 
\ee
and suppressing the subscripts denoting form degrees, one has
\be F = F^{(0)} + F'
- i \nu\Gamma \,\check J 
-\mu_0\, 
\check J\star [ 1 +i(\alpha - \beta) J]\star C'
+\mu_0 \check W^{(0)}\star D'C'\ .
\ee
where we used that $J\star D'C'\equiv 0$ being a 3-form on
${\cal Z}_2\,$ and 
${\{}V^{(0)}+V' , \check W^{(0)} {\}}_\star \equiv 0$
since $\check W^{(0)}$ is independent of $y^{\alpha}\,$.
\vspace{.3cm}

The equations of motion of \eq{Sred1b} on the above
Ansatz read
\be 
F' + g'-i\mu_0 J\star C'=0\ ,\qquad D'C'=0\ , 
\qquad \mu_0 \,(\beta - 2\alpha)\, J\star \check J_{[4]} \star C' = 0 \ , 
\ee
while plugging the Ansatz back into the action \eq{Sred1b} yields
\be 
S'_{\rm red} = i \int_{{\cal X}_4} {\check J}_{[4]} \int_{{\cal Z}_2} 
{\rm Tr}_{{\cal A}\otimes {\cal C}_4} 
\Big( 
C' \star \left[ \, \beta (F'_{[2]} + g'_{[2]}) 
- i\,\mu_0 \,\alpha \; J\star  C' \, \right] 
\Big) \ ,
\ee
from which it follows that the Ansatz leads to a
nontrivial and consistent truncation provided 
\be 
\beta = 2 \alpha\ .
\label{fixing}
\ee

In order to define the combined integration over ${\cal Z}_2$ 
and trace operation ${\cal A}$, we may take 
\be \nu=0\ ,\qquad g'=0\ .\ee
The background connection $V^{(0)}$ thereby is flat.
The simplest choice amounts to take 
$V^{(0)} = 0\,$. 
We then choose
\be {\rm Tr}_{{\cal A}\otimes {\cal C}_4}= 
\int_{{\cal Y}_2}  \frac{d^2 y}{2\pi}
{\rm Tr}_{{\cal C}_4}\ 
\ee
and make the redefinition\footnote{One could as well 
take the flat connection 
$V^{(0)} = - i z_{\alpha}\, dz^{\alpha} \,$
together with 
$s'_\alpha = - z_\alpha - 2i V'_\alpha\,$.} 
\be C'=\kappa\star b'\ ,\qquad
s'_\alpha = z_\alpha - 2i \,V'_\alpha \ ,
\ee
so that
\be F' = -\tfrac14 dz^\alpha \wedge dz^\beta \left( 
s'_\alpha \star s'_\beta + i\epsilon_{\alpha\beta} \right)\ ,
\ee
and the reduced action now reads 
\be 
S'_{\rm red} =\frac \alpha 2 \;
{\rm Vol}({\cal X}_4)\, \int_{{\cal Z}_2} d^2 z\;
{\rm Tr}_{{\cal A}\otimes {\cal C}_4} \ \kappa\star \left(is^{\prime\a}\star
s'_\a\star b' + 2b' +  i\mu_0\, b'\star b'\right)\ ,
\ee
where Vol$({\cal X}_4)=\int_{{\cal X}_4} \check J_{[4]}\,$.   
The above action is identified with the 
original action given in 
\cite{Prokushkin:1999gc} upon taking $\mu_0 = -i$ 
and Vol$({\cal X}_4) = \frac 2 \alpha\,$.

\section{Conclusions}
\label{sec:Ccl}

We have presented an action principle 
for the bosonic sector of Prokushkin and Vasiliev's 
three-dimensional matter coupled higher spin gravity. 
By integrating out the matter fields, in a fashion that
amounts to a consistent truncation in the classical
limit, followed by consistent dimensional reductions, 
we have found that the action contains various higher 
spin CS models as well as the action \cite{Prokushkin:1999gc} 
of Prokushkin, Segal and Vasiliev on twistor space.

The construction rests on three ingredients: 
i) Cartan and Vasiliev's unfolded formulation 
of classical field theory in terms of vanishing 
curvatures whereby the local degrees of freedom
arise via the Weyl zero-form (as captured
by harmonic expansions taking place in associative 
bundles independently of the 
dimension of the base manifold); ii) 
the usage of noncommutative twistor spaces for 
describing massless Weyl zero-forms in  
constantly curved backgrounds; and iii)
the AKSZ formulation of quantum field theories
based on covariant Hamiltonian actions on 
open bulk manifolds (whose boundaries contain the
classical Cartan integrable systems).

When applied to massless degrees of 
freedom in three dimensions with spin
greater than one, the above approach 
naturally leads to actions in six dimensions
containing two-dimensional noncommutative 
twistor spaces.
Their reductions on twistor space yields BF-like 
actions on four-manifolds, given by spacetimes 
times the extra auxiliary AKSZ radius, in their 
turn containing the standard symplectic 
structures of the three-dimensional 
massless gauge fields of spin greater than one 
(which are CS theories).
Thus, modulo technicalities having to do with 
consistency of the reduction schemes and
the structure of the modules making up the 
associative bundles underlying higher spin 
gravities, there is a clear overlap in four
dimensions between the standard CS formulation of 
three-dimensional higher spin gauge fields 
and the covariant Hamiltonian formulation
in six dimensions.

Turning to applications, it would be interesting to
see to what extent the action, possibly supplemented 
by boundary terms, can be used to compute the free 
energy and entropy of exact solutions of the PV system,
such as the recent nontrivial examples found in 
\cite{Iazeolla:2015tca}; for related proposals for 
on-shell actions, see \cite{Fujisawa:2013ima} and 
\cite{Apolo:2015zxh}.
The action could also have a bearing on the 
one-loop corrections from matter fields to 
the higher spin CS gauge sector. 
The above implementations may be useful in  
solidifying the Gaberdiel--Gopakumar (GG) conjecture 
\cite{Gaberdiel:2010pz}.
In particular, radiative corrections may be of 
importance in matching symmetry algebras 
\cite{Gaberdiel:2012uj} beyond the realm
of CS actions; for reviews of the CS 
approximation, see \cite{Campoleoni:2011tn},
and for existing works beyond the CS approximation, see  \cite{Chang:2011mz,Ammon:2011ua,Kessel:2015kna}.

As for alternatives to the PV system, an interesting
action for matter coupled higher spin gravity has been 
presented in \cite{Fujisawa:2013ima}.
Its four-dimensional covariant Hamiltonian 
reformulation is given by the BF-like action
\be S=\int_{{\cal X}_4} {\rm STr} \left(T\star(F+B\star \widetilde B+\tfrac12
T)-\widetilde T\star(\widetilde F+ \widetilde B\star B)+\widetilde S\star DB+
S\star \widetilde D \widetilde B\right)\ ,\label{4DFBF}\ee
where $(A,\widetilde A,B,\widetilde B;T,\widetilde T, \widetilde S, S)$
are forms of degrees $(1,1,0,2;2,2,3,1)$ valued in ${\rm Aq}^+(2)\cong hs(\tfrac12)$
and 
\be F=dA+A\star A\ ,\qquad
DB=dB+A\star B-B\star \widetilde A\ ,\ee
\be \widetilde F=d\widetilde A+\widetilde A\star \widetilde A\ ,\qquad
\widetilde D \widetilde B=
d\widetilde B+\widetilde A\star \widetilde B-\widetilde B\star A\ .\ee
The action, with its dynamical two-form $\tilde B$, 
cannot be obtained from the six-dimensional master 
action \eq{ma}, as there is an obstruction due to 
the presence of the central term $h$.
Instead, it is natural to seek a connection between 
\eq{4DFBF} and the PV system via a six-dimensional model on 
${\cal X}_4\times {\cal Z}_2$ built along the same lines as the nine-dimensional Frobenius--Chern--Simons
model in \cite{Boulanger:2015kfa}. 
The construction of such a model will be presented elsewhere. 

Comparing higher spin gravities in three and four dimensions, 
the latter admit covariant Hamiltonian actions in nine
dimensions \cite{Boulanger:2011dd,Boulanger:2015kfa}, 
though it remains unclear whether they contain the standard  
symplectic structure for Fronsdal fields \cite{Fronsdal:1978rb}
\footnote{
A related issue is whether four-dimensional
higher spin gravity contains boundary states
arising in its gauge sector, corresponding to 
an enhancement of the rigid higher spin symmetry 
algebra to the algebra of conformal higher spin 
currents.}.
Essentially, this is due to the presence of extra auxiliary 
fields in the unfolded description of Fronsdal fields
on-shell, whose inclusion into a strictly four-dimensional
off-shell formulation remains problematic.
However, the on-shell actions receive contributions as well 
from boundary terms \cite{Sezgin:2011hq,Boulanger:2012bj}
given by topological invariants that reduce on-shell 
to higher spin invariants \cite{Sezgin:2011hq}.
These invariance, which are inserted on the eight-dimensional
boundaries, are given by integrals over closed 
$p$-cycles in spacetime of on-shell (de Rham) closed 
$p$-forms in their turn given by integrals over 
twistor space of constructs built from spacetime curvatures. 
In terms of these observables, spacetime emerges in 
limits where physical states labelled by spacetime
points become separated from each other \cite{Colombo:2010fu}; 
see also the Conclusions of \cite{Iazeolla:2011cb}. 
In particular, for $p=0$, the resulting zero-form charges \cite{Sezgin:2005pv} of the (minimal bosonic) Type A and Type B models 
\cite{Vasiliev:1992av,Sezgin:2002ru} were shown in 
\cite{Colombo:2012jx,Didenko:2012tv} to provide
free theory correlation functions at the leading 
classical order, in accordance with the proposal 
made in \cite{Sezgin:2002rt,Klebanov:2002ja,Leigh:2003gk,Sezgin:2003pt}.
For boundary conditions corresponding to free fields, 
this proposal requires that correlation functions 
with separated points do not receive any radiative
corrections, in agreement with the covariant Hamiltonian 
approach \cite{Boulanger:2011dd}\footnote{Feynman diagrams 
in the nine-dimensional bulk with external zero- and one-forms 
cannot be built using only vertices containing central 
and closed forms on the twistor $Z$ space; 
see also the Conclusions of \cite{Boulanger:2012bj}.
The effects on radiative corrections from more general 
boundary conditions (including closed and central terms 
in $X$-space), nontrivial duality extensions (whereby
massless degrees of freedom are carried also by forms 
in higher degrees) and other topological effects remain 
to be seen, however.};
for a similar usage of zero-form charges 
in 3D, see \cite{Fujisawa:2013ima}.

To conclude, we view higher spin gravity as a useful 
laboratory for exploring the treatment of quantum 
field theory with local degrees of freedom by 
combining the AKSZ approach \cite{Alexandrov:1995kv} 
to topological field theories on manifolds with 
boundaries and Cartan and Vasiliev's formulation 
of nonlinear partial differential equations as 
free differential algebras with infinite zero-form 
towers.
To question its universality, it would be desirable to 
treat models in which nontrivial radiative corrections
arise in an as simple context as possible.
To this end, three-dimensional models might prove to
be fruitful and we hope that our action will be helpful 
in this endeavour.

\section*{Acknowledgements}

We thank Thomas Basile, Andrea Campoleoni, Alexander Torres Gomez,
Carlo Iazeolla, Joris Raeymaekers, Zhenya Skvortsov, 
Mauricio Valenzuela and Mikhail Vasiliev for discussions.
N.B. is F.R.S.-FNRS Research Associate and is supported in part 
by the ARC contract n$^{\rm o}$ AUWB-2010-10/15-UMONS-1.
The work of R.B. and N.B. was supported by 
the F.R.S.-FNRS (Belgium) via the  
PDR grant n$^{\rm o}$ T.1025.14 ``Gravity and Extensions''. 
P.S. is grateful to Texas A\&M University and UMONS  
for hospitality during various stages of this work.
The work of P.S. is supported by Fondecyt Regular grant N$^{\rm o}$
1140296 and Conicyt grant DPI 20140115.
The work of E.S is supported in part by NSF grant PHY-1214344.

\begin{appendix}

\section{Massive vacuum of PV system}

In this Appendix, we rewrite the massive vacuum of 
the PV system \cite{Prokushkin:1998bq} using the 
Clifford algebra variables.
The vacuum solution reads
\be B^{(0)} = \nu\,\Gamma\ ,\quad A^{(0)}= W^{(0)}+ V^{(0)}\ ,\quad \nu\in \Real\ ,\quad
\Gamma=\Gamma^\dagger \in {\cal C}_4^+\ ,\ee
obeying
\begin{equation}
d_X W^{(0)}+W^{(0)}\star W^{(0)}=0\;,\qquad d_X S^{(0)}_\a+[W^{(0)},S^{(0)}_\a]_\star=0\ ,
\label{vac1}\ee
\be
[\Gamma, W^{(0)}]=0\;,\qquad [\C,S^{(0)}_\a]_\star=0\;,\qquad [S^{(0)}_\a,S^{(0)}_\b]_\star=-2i\,\e_{\a\b}\big(1-\nu\,\C\k\big)\;,\label{vac2}
\end{equation}
where $S^{(0)}_{\a}=z_\a-2i V^{(0)}_\a$.
The constraints along ${\cal M}_3$ can be solved using a 
gauge function that commutes to $\Gamma$, \emph{viz.}
\be W^{(0)}=L^{-1}\star dL\ ,\qquad S^{(0)}_\a= L^{-1}\star \tilde z_\a\star L
\ ,\qquad [\Gamma,L]_\star =0\ ,\ee
where $\tilde z_\a$ obeys
\be d_X \tilde z_\a=0\ ,\qquad [\Gamma, \tilde z_\a]_\star =0\qquad
[\tilde z_\a,\tilde z_\b]_\star=-2i\,\e_{\a\b}\big(1-\nu\,\C\k\big)\ .
\label{rc}\ee
The integrability on ${\cal Z}_2$ implies the existence
of a double $\tilde y_\a$ obeying 
\be [\yt_\a,\zt_\b]_\star=0\ ,\qquad \{\Gamma,\yt_\a\}=0\ ,\qquad \tilde y_\a\rvert_{\nu=0}
=A y_\a\ ,\qquad \{A,\Gamma\}=0\ ,\label{rc3}\ee
for some matrix $A\in{\cal C}_4\,$. 
Thus one can take 
\be L=L(\tilde y_\a,\Gamma_{ij})\ ,\qquad S^{(0)}_\a= \tilde z_\a\ .\ee
Remarkably, as found in \cite{Prokushkin:1998bq}, the solution obeys 
\begin{equation}
[\yt_\a,\yt_\b]_\star=2i\e_{\a\b}\big(1-\nu\,\C\big)\;.\label{rc2}
\end{equation}
At the level of the complexified algebra, a solution is given by
\begin{equation}
\zt^\Comp_\a=X\,z_\a+Y\,\sigma_\a\;,\quad \yt^\Comp_\a=A\,y_\a+B\,\tau_\a\;,
\end{equation}
where $X$, $Y$, $A$ and $B$ are built out of gamma matrices $\Gamma_i$
and the building blocks
\begin{equation}
\sigma_\a:=\nu\int_0^1dt\,t\,e^{ityz}(y_\a+z_\a)\;,\qquad \tau_\a:=\nu\int_0^1dt\,(t-1)\,e^{ityz}(y_\a+z_\a)\;,
\end{equation}
with $yz:=y^\a z_\a$, obey
\begin{equation}
\begin{split}
&[z_{[\a},\sigma_{\b]}]_\star=-i\nu\,\e_{\a\b}\,\k\;,\quad [\sigma_\a,\sigma_\b]_\star=0\;,\quad \{\sigma_\a,\tau_\b\}_\star=0\;,\\
&[z_\a,\tau_\b]_\star=\{\sigma_\a,y_\b\}_\star\;,\quad \{y_{[\a},\tau_{\b]}\}_\star=i\nu\,\e_{\a\b}\;,\quad [\tau_\a,\tau_\b]_\star=0\;.
\end{split}
\end{equation}
For the above Ansatz, Eqs. \eq{rc}, \eq{rc3} and \eq{rc2}, respectively, are equivalent to
\be
X^2=1\;,\quad XY=YX=-\C\;,\quad [\Gamma,X]=0=[\Gamma,Y]\ ,\ee
\be [A,X]=\{A,Y\}=[B,X]=\{B,Y\}=0\;,\quad \{\Gamma,X\}=0=\{\Gamma,Y\}\ ,
\ee
\be
 A^2=1\ ,\qquad AB=-BA=-\Gamma\;.
\ee
A solution for $\tilde z_\a$ that commutes to ${\cal C}_4^+$ 
is obtained by taking \cite{Prokushkin:1998bq}\footnote{
Using the gauge function, the first order fluctuation $B^{(1)}$
can be written as $B^{(1)}= L^{-1}\star B^{\prime(1)}\star L$
where $d_XB^{\prime(1)}=0$ and $[\tilde z^\a,B^{\prime(1)}]_\star=0$.
Thus, if $\tilde z^\a$ commutes to ${\cal C}_4^+$ then 
$B^{\prime(1)}=B^{\prime(1)}( \tilde y^\a, \Gamma_{ij})$.
Otherwise, as $\tilde z_\a=z_\a-2i V^{\prime(0)}_\a$ where
$V^{\prime(0)}=L\star (d_Z+V^{(0)})\star L^{-1}$, one may 
equivalently solve $d_Z B^{\prime(1)} + 
[V^{\prime(0)}, B^{\prime(1)}]_\star =0$ iteratively using a homotopy contractor in ${\cal Z}_2$-space.}
\be
\C=\C_{1234}\ ,\quad X=1\;,\quad Y=-\C\;,\quad A=\C_1\;,\quad B=-\C_{234}=-\C_1\C\;,
\ee
where $\C_{i_1..i_k}:=\C_{[i_1}..\C_{i_k]}$, and the relation 
between the PV generators and ours is given in \eq{map1}, which 
yields
\be
\zt_\a=z_\a-\nu(y_\a+z_\a)\int_0^1 dt\,t\,e^{it\,yz}\,\C\;,\quad \yt_\a=\C_1\,\Big[y_\a-\nu(y_\a+z_\a)\int_0^1 dt\,(t-1)\,e^{it\,yz}\,\C\Big]\;,
\ee
This solution, however, does not satisfy the required reality conditions,
\emph{i.e.} $((\tilde y_\a^\Comp)^\dagger,(\tilde z_\a^\Comp)^\dagger)$
form a set of deformed oscillators that is linearly independent from
$(\tilde y_\a^\Comp,\tilde z_\a^\Comp)$
This is remedied by a highly nontrivial modification found in 
\cite{Prokushkin:1998bq}  given by
\be
\begin{split}
\zt^{\rm sym}_\a &= z_\a+\frac{\nu}{8}\,\int_{-1}^1ds(1-s)
\Big[e^{\frac{i}{2}(s+1)yz}(y_\a+z_\a)\star
\Phi(\tfrac12,2;-\C\k\,\ln\abs{s}^{-\nu})
\\
&+e^{\frac{i}{2}(s+1)yz}(y_\a-z_\a)\star
\Phi(\tfrac12,2;\C\k\,\ln\abs{s}^{-\nu})\Big]\star\k\,\C
\\[2mm]
\yt^{\rm sym}_\a &= \C_1\Big[y_\a+\C\,
\frac{\nu}{8}\,\int_{-1}^1ds(1-s)\,e^{\frac{i}{2}(s+1)yz}
\Big((y_\a+z_\a)\,\Phi(\tfrac12,2;-\C\,\ln\abs{s}^{-\nu})
\\
&-(y_\a-z_\a)\,\Phi(\tfrac12,2;\C\,\ln\abs{s}^{-\nu})\Big)\Big]\;,
\end{split}
\ee
where $\Phi(a,b;z)$ is the confluent hypergeometric function.
\end{appendix}

\section{BF-like formulation of modified Blencowe action}

In this Appendix, modulo a technical assumption on a normalization
coefficient (given in Eq. \eq{calN}), we present a consistent 
truncation of the master action \eq{ma} for nontrivial vacuum 
expectation value of $B$ leading to a model on ${\cal X}_4$ 
in which the gauge fields and the Lagrange multipliers belong 
non-isomorphic dual spaces.

To this end, we observe that the equations of motion \eq{e4} 
admit a consistent truncation given by 
\be B=\nu\Gamma\in \Real\ ,\label{Bnu}\ee
as can be seen from the fact that the resulting field equations, 
\be 
F + \nu\Gamma h+g- h\star \mu\star T= 0 \ ,\quad  DT = 0 \ ,\quad 
DS +h\star T+ \mu \star S\star S=0\ ,
\ee
form a Cartan integrable system.
Inserting \eq{Bnu} into the master action \eq{ma}, the resulting consistently truncated 
action is given by 
\be 
S_{\rm red}[A;T]=\int_{{\cal M}_6} {\rm Tr}_{{\cal A}\otimes {\cal C}_4}
T\star (F+\nu\Gamma h+ g-\tfrac12 h\star \mu \star T)\ .
\label{rednu}
\ee
Its gauge symmetries take the form
\begin{equation}
\delta A=D\e+\mu\star h\star\eta\;,\quad\delta T
=D\eta-[\e,T]_\star\;,
\end{equation}
and the equations of motion and boundary conditions are given by
\begin{equation}
F+\nu\Gamma h + g-\mu\star h\star T=0\;,\quad DT=0\;,\quad T\rvert_{\partial{\cal M}_6}=0\;.
\end{equation}
To reach a Blencowe type action, we assume Eq. \eq{reduceg} and make the reduction
\be A = V^{(0)}_{[1]}+\widetilde{W}_{[1]} - \widetilde K_{[1]}-\mu_0 J \star 
\widetilde{K}_{[1]}\ ,\ee
\be  
T = \widetilde{T}_{[2]}+ \widetilde K_{[1]}\star \widetilde K_{[1]} -
\mu_0 J\star \widetilde{T}_{[2]}\;,
\ee
where the subscripts indicating the form degree will be suppressed
henceforth, and $V^{(0)}$ is a twistor space background connection obeying
\be dV^{(0)}+ V^{(0)}\star V+ \nu\Gamma J=0\ ,
\qquad V^{(0)}\rvert_{\nu=0}=0\ ,\ee
and the reduced fields 
\be \tilde{f}\in \Omega({\cal X}_4) \otimes \widetilde {\cal A}\otimes 
\widetilde{\cal C}_4 \ ,\quad 
\label{widetilde}\ee
where $\widetilde {\cal A}$ is an associative algebra 
generated by the deformed oscillator $\tilde y^\a$ 
obeying
\be d\tilde y^\a+[V^{(0)},\tilde y^\a]_\star =0\ ,\qquad 
[\tilde y^\a,\tilde y^\b]_\star = 2i\e^{\a\b}(1-\nu \Gamma)\ ,\qquad 
\tilde y_\a\rvert_{\nu=0}
=\Gamma_1 y_\a\ ;\label{defoscalg}\ee
see the Appendix A for further details.
Thus, 
\be \pi(\tilde f)=\tilde f\ ,\qquad d\tilde{f}+[V^{(0)},\tilde{f}]_\star=d_X\tilde{f}\ ,\ee
and the reduced equations of motion and boundary conditions
are given by the counterparts of Eqs. \eq{red1}--\eq{red3} with all
quantities now valued in \eq{widetilde}, which form a 
Cartan integrable system.
The consistently reduced action reads
\be
\widetilde S_{\rm red}[\widetilde W,\widetilde T]=-\mu_0 \int_{{\cal X}_4} 
\int_{{\cal Z}_2} {\rm Tr}_{{\cal A}\otimes{\cal C}_4}
J\star \widetilde {T}\left( \widetilde{F}+\check g +\check g '+
\tfrac12 \widetilde{T}\right)\ .\ee

To obtain the alternative model, we make the choice 
\be 
{\rm Tr}_{{\cal A}\otimes {\cal C}_4}=\int_{{\cal Y}_2}   
\frac{d^2y}{2\pi} {\rm Tr}_{{\cal C}_4} \Gamma\ ,
\label{choice2}
\ee 
corresponding to $m=1$ in Section 4.1.
We also take
\be 
\widetilde W\in \Omega_{[1]}({\cal X}_4) \otimes {\rm Aq}^+(2;\nu)\otimes 
{\cal C}_4^+\ ,
\label{Wdef}
\ee
\be 
\widetilde T\in \Omega_{[2]}({\cal X}_4) \otimes \rho({\rm End}^+({\cal F}^\s_\nu))\otimes 
{\cal C}_4^+\ ,
\label{Tdef}
\ee
where ${\cal F}^\sigma_\nu$ is the Fock representation 
space of ${\rm Aq}(2;\nu)$ with ground state having 
eigenvalue $\sigma$ of the Klein operator $-\Gamma$, and
\be \rho:{\rm End}({\cal F}^\sigma_\nu)\rightarrow
{\rm Aq}(2;\nu)\ee
is a monomorphism given by the deformed oscillator realization of 
${\rm End}({\cal F}^\sigma_\nu)$.
The space ${\rm End}^+({\cal F}^\s_\nu)$ consists of the endomorphisms
that commute to $\Gamma$.
Its oscillator realization
\be \rho({\rm End}^+({\cal F}^\s_\nu)) =\bigoplus_{\s'=\pm}
{\rm Aq}^{\s'}(2;\nu)\star P^{\s}_\nu \star {\rm Aq}^{\s'}(2;\nu)\ ,\ee
where 
\be {\rm Aq}^\s(2;\nu)=\bigoplus_{\s'=\pm} \Pi^{\s'}_\Gamma\star {\rm Aq}(2;\nu)
\star \Pi^{\s\s'}_\Gamma\ ,\ee
and 
\be P^\s_\nu= \frac{2}{1+\nu \sigma}\Pi^{-\s}_\Gamma \star  \left[{}_1F_1(\tfrac32; \tfrac{3+\nu\sigma}2;-2w)\right]^{\rm W}\ ,\ee
where 
\be w=\{a^+, a^-\}_\star\ ,\qquad a^\pm =u^\pm_\a y^\a\ ,\quad u^{-\a} u^+_\a=\tfrac i2\ ,\ee
is the symbol of the oscillator realization of the ground state projector 
in ${\rm End}({\cal F}^\s_\nu)$ given in Weyl order; for details,
see \cite{Boulanger:2013naa,Boulanger:2015uha}.
The reduced action reads
\be \check S_{{\rm II}_\nu}= -\mu_0  \int_{{\cal X}_4} \int_{{\cal Z}_2\times {\cal Y}_2}
\frac{d^2y}{2\pi}\,{\rm Tr}_{{\cal C}_4} \Gamma J\star 
\widetilde T\star (\widetilde F+g+g'+\tfrac12 \widetilde T)\ ,\ee
where $\widetilde T\star (\widetilde F+g+g')$ and $\widetilde T\star \widetilde T$ lie 
in $\rho({\rm End}^+({\cal F}_\nu))$.
The Lagrangian is finite provided that 
\be 
{\cal N}^\s_\nu=\int_{{\cal Z}_2\times {\cal Y}_2}\frac{d^2y}{2\pi}
\,{\rm Tr}_{{\cal C}_4} \Gamma J\star P^\s_\nu\ ,
\label{calN}
\ee
is finite.
If so, the dual pairing displayed in \eq{Wdef} 
and \eq{Tdef} is non-degenerate, and 
the equations of motion and boundary 
conditions read
\be \widetilde F+g+ g'+ \widetilde T\approx 0\ ,\quad \widetilde D \widetilde T\approx 0\ ,
\qquad \widetilde T\rvert_{\partial{\cal X}_4}=0\ ,\ee
where the first equation is valued in ${\rm Aq}^+(2;\nu)$
and the second equation in $\rho({\rm End}^+({\cal F}_\nu))$.
Eliminating $\check T$ via the first equation (by
inverting the monomorphism), yields
\be \check S_{{\rm II}_\nu}\approx \frac12\mu_0  \int_{{\cal X}_4} \int_{{\cal Z}_2\times {\cal Y}_2}
\frac{d^2y}{2\pi}\,{\rm Tr}_{{\cal C}_4} \Gamma J\star 
(\widetilde F+g+g')\star (\widetilde F+g+g')\ ,\ee
which is formally divergent for gauge fields given by finite
polynomials, unless 
\be \widetilde F+g+ g'\approx 0\ .\ee

It is possible to construct a deformation of Blencowe's action
by making use of Vasiliev´s (graded cyclic) supertrace operation ${\rm STr}_\nu$
on the Weyl algebra ${\rm Aq}(2;\nu)$ of the deformed oscillator 
algebra \eq{defoscalg} \cite{Vasiliev:1989re}, which is uniquely
characterized by ${\rm STr}_\nu 1=1$ and ${\rm STr}_\nu \Gamma=\nu$ 
(and hence differs from the trace operation proposed above).
Using this operation, it straightforward to deform
Blencowe's action in ${\cal X}_3$ and uplift it 
to BF-type model in ${\cal X}_4$ with action
\be \check S_{{\rm I}_\nu}= -\mu_0  \int_{{\cal X}_4} {\rm STr}_{\nu}
\,{\rm Tr}_{{\cal C}^+_4} \Gamma \, 
\widetilde T\star (\widetilde F+g+g'+\tfrac12 \widetilde T)\ ,\ee
where $\widetilde T$ and $\widetilde W$ are valued 
in ${\rm Aq}^+(2;\nu)\otimes {\cal C}_4^+$.

Whether there exists a modification of \eq{choice2} 
that yields the ${\rm STr}_\nu$ operation starting from
the master action in six dimensions, possibly by using
the trace operation \eq{choice} for $m=0$, remains to be seen\footnote{
The operation $\int_{{\cal Y}_2} {\rm Tr}_{{\cal C}_4}$ is formally
graded cyclic on ${\rm Aq}(2;\nu)$, but since it sends $1$ and 
$\Gamma$ to $1$ and $0$, respectively, it cannot be proportional
to ${\rm STr}_\nu$ and hence it must be ill-defined.}.

Thus, provided that ${\cal N}^\s_\nu$ is finite, 
we have found a covariant Hamiltonian action in 
which the gauge fields and the Lagrange multipliers 
belong non-isomorphic dual spaces that is 
an alternative to $\nu$-deformed Blencowe's action.
Whether this action admits a coupling to matter
is unknown, while the  action presented above 
appears to be amenable to such couplings on ${\cal M}$.


\providecommand{\href}[2]{#2}\begingroup\raggedright\endgroup

\end{document}